# Predicting Young's Modulus of Glasses with Sparse Datasets using Machine Learning

*Suresh Bishnoi[1,#], Sourabh Singh[1,#], R. Ravinder[1], Mathieu Bauchy[2], Nitya Nand Gosvami[3]*
*Hariprasad Kodamana[4,\*], N. M. Anoop Krishnan[1,3,\*]*

[1]Department of Civil Engineering, Indian Institute of Technology Delhi, Hauz Khas, New Delhi 110016, India

[2]Physics of AmoRphous and Inorganic Solids Laboratory (PARISlab), Department of Civil and Environmental Engineering, University of California, Los Angeles, CA 90095, USA

[3]Department of Materials Science and Engineering, Indian Institute of Technology Delhi, Hauz Khas, New Delhi 110016, India

[4]Department of Chemical Engineering, Indian Institute of Technology Delhi, Hauz Khas, New Delhi 110016, India

[\*]Corresponding authors: H. Kodamana (kodamana@iitd.ac.in), N. M. A. Krishnan (krishnan@iitd.ac.in)
[#]Both the authors contributed equally.

**Abstract**

Machine learning (ML) methods are becoming popular tools for the prediction and design of novel materials. In particular, neural network (NN) is a promising ML method, which can be used to identify hidden trends in the data. However, these methods rely on large datasets and often exhibit overfitting when used with sparse dataset. Further, assessing the uncertainty in predictions for a new dataset or an extrapolation of the present dataset is challenging. Herein, using Gaussian process regression (GPR), we predict Young's modulus for silicate glasses having sparse dataset. We show that GPR significantly outperforms NN for sparse dataset, while ensuring no overfitting. Further, thanks to the nonparametric nature, GPR provides quantitative bounds for the reliability of predictions while extrapolating. Overall, GPR presents an advanced ML methodology for accelerating the development of novel functional materials such as glasses.

*Keywords: Neural network, Gaussian process regression, Silicate glasses, Young's modulus, Sparse dataset*

**Introduction**

Glasses are ubiquitously used for a wide-range of applications such as smart phone screens, optical fibers, wind shields, and even for nuclear waste immobilization[1]. In order to address the ever increasing infrastructural and energy requirements, discovery of novel glass compositions with properties tailored for particular applications is required[1,2]. Predicting the composition–property relationships holds the key to development of such novel compositions. However, developing this map is an extremely challenging task in glasses due to the following reasons. (i) Glasses can be formed of virtually any element or its oxide, provided the structure is cooled fast-enough from the liquid state to avoid crystallization. This allows formation of glasses with any stoichiometry, thereby making the possible glass compositions nearly infinite[3,4]. (ii) Silicate glasses exhibit highly complex and non-linear

composition–property relationships preventing any direct extrapolation from a few compositions[3]. As such, developing physics-based models for property predictions in glasses is still an open challenge that needs to be addressed.

An alternate approach to predict new materials is to use data-based modeling techniques such as machine learning[2,5–9]. These methods rely on available data, either from simulations or experiments, to develop models that capture the hidden trends in the input–output relationships. One of the widely used and attractive techniques in ML is neural network (NN)[10–12]. NN is a method, inspired from the neurons in the brain, wherein a non-linear network of hidden layer units "learn" from the data. NN has been successfully used to address a wide range of problems exhibiting highly non-convex and nonlinear input-output relationships[2,5,6,9,13–16]. In particular, ML has been successfully used in oxide glasses to predict a wide range of equilibrium and nonequilibrium composition–property relationships such as liquidus temperature[17], solubility[18], glass transition temperature[19], stiffness[20], and dissolution kinetics[21].

Despite wide-spread applications, NN-based methods have a few inherent deficiencies that makes it unfavorable for material informatics. Being a parametric method, NN relies on the availability of large-scale data for reliable training[19,22,23]. However, obtaining such consistent large datasets require a large number of experiments or numerical simulations that are prohibitive, if not significantly expensive. For example, in the case of glasses, samples are produced by the traditional melt-quench process following which further experiments, such as nanoindentation, are required to measure the Young's modulus. Carrying out such experiments on a large sample set or compositional space for would be nearly impossible. On the other hand, sparse datasets that are consistent and accurate can be obtained even at the laboratory scale or from physics-based simulations. However, these sparse datasets may pose some unique problems for NN-based ML algorithms as follows. (i) Development of reliable weights in NN requires training over a sizeable data that is coming from a consistent dataset. While experiments or simulations may provide consistent data, such datasets may be limited in size, hence making it challenging to develop reliably trained networks[11]. (ii) Despite its ability to infer hidden trends within the dataset, NN can exhibit overfitting due to the parametric nature of the method. Overfitting suggests that the noise of the data is memorized instead of identifying the underlying trend[5,24,25], which can occur in a parametric method such as NN. This situation is highly undesirable as it reduces the capability of NN to predict untrained data coming from the same dataset. (iii) Finally, obtaining the uncertainty of predictions of a trained NN on a new dataset is challenging. This makes it unreliable to apply NN for untrained compositions, or even to extrapolate from the trained dataset.

These deficiencies can be addressed by using an advanced nonparametric machine learning algorithm, namely, Gaussian process regression (GPR)[26,27], which uses a probabilistic framework for predictions[28,29]. In GPR, the prior dataset coming from experiments or simulations is assumed to be coming from an underlying Gaussian distribution with a well-defined mean and standard deviation. Thus, the objective of GPR is to estimate the underlying normal distribution by minimizing the error in the prior. Once the function is obtained through regression, the interpolation for any other input variables can be obtained from resulting distribution. Note that since the predictions in GPR are from a distribution, the predicted value corresponds to the mean value and confidence intervals for the prediction are provided by the standard deviations. Due to this unique feature, GPR presents a robust machine learning methodology to develop reliable and accurate predictive model for material informatics[7,30].

Herein, we present a machine learning methodology using Gaussian process regression (GPR) that can "learn" composition–property relationships from sparse datasets. Using a few sparse datasets of Young's modulus for various silicate glasses, we show that GPR can outperform widely used machine learning techniques such as NN. Additionally, GPR provides error bounds for the predicted values thereby providing a quantitative estimate for the reliability of the prediction. Overall, we show that GPR presents a robust and transferable technique for material informatics, that can be used to develop novel materials using even a sparse dataset.

**METHODOLOGY**
**Data set**
The datasets used in the prediction comprise the Young's modulus values for four different families of silicate glasses along with their compositions and density. The input data set includes the molar percentage composition of the oxide components and the density of the glass. Output data set is the Young's Modulus of the glass compositions. The values of these glasses are obtained from the INTERGLAD© Ver.7 international glass database[32]. The four glass families considered herein are – (a) calcium aluminosilicate (CAS) used as alkali-free display glasses, (b) sodium calcium silicate used as archetypical window glasses (NCS), (c) sodium germanium silicate (NGS), and (d) sodium borosilicate (NBS) used for nuclear waste immobilization. Table 1 shows the relevant features of these glasses including the elements serving as network former and network modifier along with the coordination numbers of the network formers and the dataset size.

Note that the glasses are chosen here are so as to represent different classes of silicate glasses exhibiting distinct features as follows. (i) CAS presents a glass having two network formers (Si and Al) and a network modifier (Ca)[33]. Al exhibits a tetrahedral structure exhibiting a net negative charge which is charge balanced by a $Ca^{2+}$ cation in the vicinity that does not form a non-bridging oxygen (NBO). Further, Al preferentially bonds with Si rather than Al in accordance with the Loewenstein rule[34]. (ii) NCS presents a glass having two network modifiers (Ca and Na) and a network former (Si). Here, both the network modifiers create NBO with Ca creating more NBOs than Na, which can have some significant effects on the mechanical properties such as hardness[35,36]. (iii) NGS presents a glass having two network formers (Si and Ge) and a network modifier (Na). Here, the network modifier creates an NBO, while both the network formers form tetrahedral structure with O atoms. (iv) NBS presents a glass having two network formers (B and Si), and a network modifier (Na). Here, depending on the percentage of Na, B can have a coordination number of three or four[37]. Note that this differential coordination number of B results in a non-monotonic evolution of mechanical properties. Overall, each of these glasses present a unique structure depending on the local and global composition, thereby exhibiting a complex composition–structure relationship.

| Table 1: Silicate glasses considered herein along with their network former, network modifier species, coordination number, and datasize ||||
|---|---|---|---|
| **Glass** | **Network former** | **Network modifier** | **Coordination number (Network former)** | **Data size** |

| Calcium aluminosilicate (CAS) | Silicon (Si), Aluminum (Al) | Calcium (Ca) | 4 (Al), 4(Si) | 42 |
| Sodium calcium silicate (NCS) | Silicon (Si) | Sodium (Na), Calcium (Ca) | 4 (Si) | 46 |
| Sodium germanosilicate (NGS) | Silicon (Si), Germanium (Ge) | Sodium (Na) | 4 (Si), 4 (Ge) | 46 |
| Sodium borosilicate (NBS) | Silicon (Si), Boron (B) | Sodium (Na) | 4 or 3 (B), 4 (Si) | 105 |

The datasize for each of these glasses are given in Table 1. The dataset size is limited by the availability of relevant experimental data for the considered series of glasses. Further, Table 2 presents the range of values of compositions for each of the oxide components. It should be noted that the compositions are chosen in such a way that the sum of mole percentage of individual oxide components add up to 100%. This is to ensure that the compositions considered herein are pure and does not have any additional "noise" due to small variations in data.

| Table 2: Glass compositions considered herein with the range of each of the individual oxide components | | | | |
|---|---|---|---|---|
| **Glass** | **Composition** | **x (mol %)** | **y (mol %)** | **1-x-y (mol %)** |
| CAS | $(CaO)_x(Al_2O_3)_y(SiO_2)_{1-x-y}$ | 4.00 to 68.27 | 2.90 to 36.10 | 16.41 to 66.00 |
| NCS | $(Na_2O)_x(CaO)_y(SiO_2)_{1-x-y}$ | 4.00 to 27.50 | 60.00 to 82.00 | 4.50 to 23.80 |
| NGS | $(Na_2O)_x(GeO_2)_y(SiO_2)_{1-x-y}$ | 3.38 to 33.33 | 6.68 to 48.31 | 33.33 to 86.64 |
| NBS | $(Na_2O)_x(B_2O_3)_y(SiO_2)_{1-x-y}$ | 10.00 to 96.91 | 2.60 to 80 | 0.49 to 39.39 |

**Machine learning algorithms**

We employ various supervised learning models to model the data. The available data set that comprises of inputs and outputs is randomly divided into (i) a training set and (ii) a test set. While doing so, it is ensured that the training and test data sets have reasonable spread over each variables' span. The training set is first used to train the model, that is, to optimize the parameters that relate the inputs to the outputs. The test set, which is fully unknown to the model, is then used to assess the performance of the model—by comparing the outcomes of the model, for inputs which it has not been explicitly trained for, to the true reference outputs. Here, 70% and 30% of the data are attributed to the training and test sets, respectively. Note that, in the case of the NN method, the training data is further divided into 55% as training set and 15% as validation set, totaling to 70% of the data. In the following, we provide a brief description of the various supervised learning models used herein.

*(i) Neural network (NN)*

NN is a nonlinear supervised learning model that has immense capabilities to capture complex data trends[11,12]. A NN consists of the input, hidden, and output layers, wherein the hidden layer contains a given number of units that take their inputs from the input layer and connect to the output layer. A weight is attributed to the links that connects two units. The output ($h_i$) of a hidden layer unit $i$ is calculated as:

$$h_i = s\left(\sum_{j=1}^{p} V_i x_i + T_i^{hid}\right)$$ Equation (1)

where $s()$ is the activation function (or transfer function), $N$ the number of inputs, $V_i$ the weights of $i^{th}$ layer, $x_i$ the input values, and $T_i^{hid}$ is the bias term. To account for the non-linearity in the composition–Young's modulus relationship, the activation function used herein is a sigmoid as given below [10,38,39]:

$$s(u) = \frac{1}{1+e^{-u}}$$ Equation (2)

The network is first trained with the training data to obtain the weight parameters between input, hidden layer and output layer units. Finally, the predictive capability of the network is evaluated using the data from the test set.

*(ii)    Gaussian Process Regression (GPR)*

The GPR modeling paradigm tries to find a distribution over a set of possible nonparametric functions for representing the relationship between a set of input and output datasets[26,27]. Traditionally, this relationship is characterized with various classes of parametric functions that have a fixed model structure as in the case of OLS or NN. Given a training set $\{(x_1, y_1), (x_2, y_2), \ldots, (x_n, y_n)\}$, a GPR model explains the response $y_i$ introducing latent variables, $f(x_i), i = 1, 2, 3, \ldots, n$ from a Gaussian process (GP), and explicit basis functions, $g(.)$. Here, the GP is the set of random variables, $f(x_i), i = 1, 2, 3, \ldots, n$, such that they have a joint Gaussian distribution having mean function $m(x_i)$ and covariance function, also known as kernel function, $k(x_i, x') = Cov(x_i, x')$. Given, $f(x_i)$ and $x_i$, the regression output $y_i$ is obtained from the following probability distribution

$$P(y_i|f(x_i), x_i) \sim N(y_i|g(x_i)^T \beta + f(x_i), \sigma^2)$$ Equation (3)

where $(x) \sim GP(0, k(x, x'))$, $g(x)$ are a set of nonlinear basis functions that transform the original feature vector $x$ and $\epsilon \sim N(0, \sigma^2)$ is the noise term corresponding to the signal noise. As the latent variable $g(x_i)$ is introduced for each observation $x_i$, the GPR model is not having fixed functional form, rendering it to be nonparametric.

Here, we use two popular kernel functions, namely, exponential function and automatic relevance determination (ARD) exponential function, for analyzing the performance of the GPR model. The exponential kernel has the following form

$$k(x, x') = \frac{|x-x'|}{l}$$ Equation (4)

where $l$ is the length scale parameter, which is fixed. While for ARD exponential, the length scale is parameter is continuously leaned over time, aiding flexibility. The length scale parameters determine the relevancy of input data to the regression and it is tuned to be larger if an input is irrelevant for regression output.

**Makishima Mackenzie model**

The Makishima-Mackenzie (MM) model[40] can be used to predict the Young's modulus of oxide glasses from their chemical composition and the density. This method is derived from the idea that the elastic energy is proportional to the dissociation energy of the oxide constituents per unit volume along with their packing efficiency. For multicomponent glasses, the Young's modulus is given by

$$Y = 2V_t \sum_j G_j X_j$$ Equation (5)

where $G_i$ and $X_i$ are the dissociation energy per unit volume and mole fraction of the component $j$, respectively. The packing density $V_t$ of a glass with density $\rho$ given by

$$V_t = \frac{\rho}{M}\sum_j V_j X_j \qquad \text{Equation (6)}$$

where $M$ is the molecular weight, $\rho$ is the density, and $V_j$ is a packing factor. For a single component oxide glass of composition $A_{\alpha_j}O_{\beta_j}$, $V_j$ is:

$$V_j = 6.023 \times 10^{23} \frac{4}{3}\pi(\alpha_j R_A^3 + \beta_j R_O^3) \qquad \text{Equation (7)}$$

where $R_A$ and $R_O$ are the respective ionic radii of the cation and oxygen. Combining Equations (5), (6), and (7), Young's modulus can be expressed as:

$$Y = \frac{\rho}{M}\sum_j \left(6.023 \times 10^{23} \frac{4}{3}\pi(\alpha_j R_A^3 + \beta_j R_O^3)\right) X_j \times \left(\sum_j G_j X_j\right) \qquad \text{Equation (8)}$$

## RESULTS

### Neural Network

We first focus on the predictions obtained using NN (see Methodology). Note that the accuracy of predictions in NN can be improved by optimizing the number of hidden layer units. To this extent, we train the dataset against NN with varying number of hidden layer units. Figure 1(a) shows the variation of $R^2$ values for training and test sets with respect to the number of hidden layer units in the NN. We observe an increase in the performance of the NN during training with respect to the increase in the number of hidden layer units, as evident from the $R^2$(training) which increases with increasing hidden layer units. In the case of test set, we observe that the $R^2$(test) initially increases with the number of hidden layer units. This suggests that the predictive capability of the NN is increasing with increasing hidden layer units. However, beyond a critical number of hidden layer units, six in this case (see Fig. 1(a)), we observe that the $R^2$(test) starts decreasing drastically. This is due to overfitting wherein the parameters of the NN capture the noise along with the underlying relationship in the training dataset. As such, any input other than that from the training set will result in a poor prediction by this NN. Thus, the optimum value of the number of hidden layer units corresponds to that wherein the $R^2$(test) exhibits a maximum or near maximum, while ensuring that the training set also exhibits a near maximum $R^2$ value. Based on this analysis, we chose a net with six hidden layer units to predict the Young's modulus of CAS glasses.

Figure 1(b) shows the predicted values of Young's modulus for the CAS glasses with respect to the true measured values for a NN with six hidden layer units. To benchmark the performance of NN, we also performed ordinary least squares (OLS). We observe that while the results for the training set is notably improved for NN with an $R^2$(training) of 0.772 in comparison to the OLS method with an $R^2$(training) of 0.694, the test set exhibits a notably poorer performance ($R^2$(test) = 0.590 for NN against $R^2$(test) = 0.693 for OLS). This suggests that even the optimized NN exhibits lower poorer predictive capability than the OLS for small dataset. This exemplifies the inherent limitation of NN, which requires a large number of data points in the training set to develop a reliable network with appropriate weights corresponding to each unit.

In order to ensure the generality of our results, we extend our study to three more silicate glass compositions—namely, sodium calcium silicate (NCS), sodium germanium silicate (NGS), and sodium borosilicate (NBS)—with varying dataset size (see Methodology). Note these glasses are chosen so as to represent different features such as multiple network formers, or network modifiers, or coordination numbers (see Methodology). Further, the composition–structure relationships in these glasses are highly non-trivial and depends closely on the local environment around each atomic species. As such, the ability to predict the Young's modulus from the compositions of these glasses provides an insight into the transferability and robustness of ML algorithms.

Figures 2(a)–(c) show the predicted values of Young's modulus with respect to measured values for NCS, NGS, and NBS glasses using NN, respectively. We observe that the prediction is improved when the size of the dataset is increased. Figures 2(d)–(f) shows the variation of $R^2$ values for training and test set with respect to number of hidden layer units for NCS, NGS, and NBS glasses, respectively. As in the case, of CAS glasses, we observe overfitting with increasing number of hidden layer units. Further, for the optimum number of hidden layer units, we observe that the results are significantly improved when the training set is larger. This could be attributed to the fact that NN requires a large training set to ensure reliable prediction without overfitting. As such, the small size of data obtained from experiments (as in the present case) or simulations would be insufficient to train a robust NN, reducing its utility and rendering it unreliable.

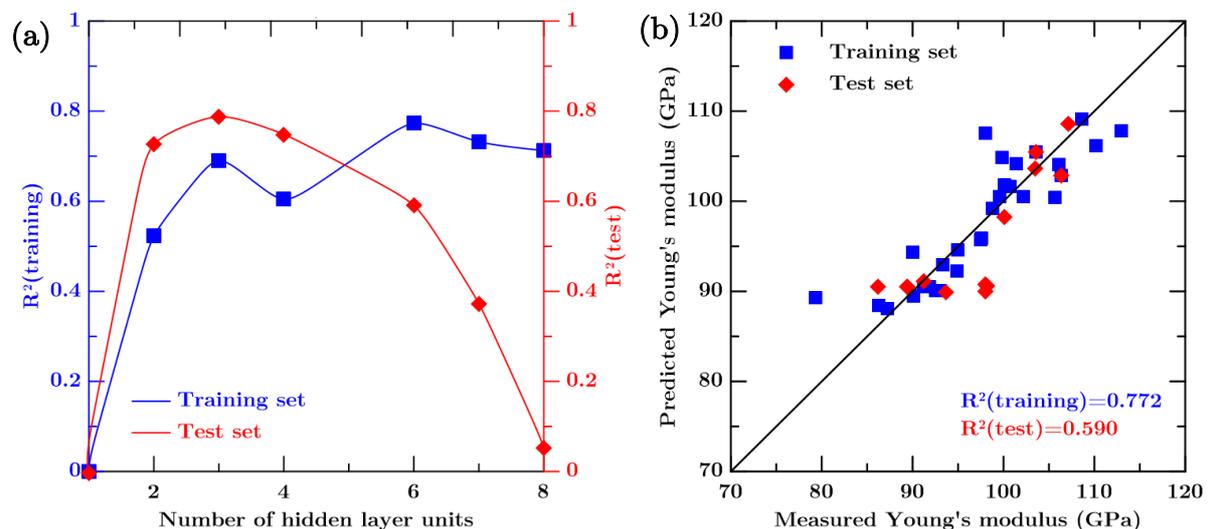

**Figure 1. (a)** $R^2$ of training set (left axis) and $R^2$ of the test set (right axis) with respect to the number of hidden layer units in the NN for calcium aluminosilicate (CAS) glasses. Lines are guide for eye. **(b)** Predicted Young's modulus with respect to measured Young's modulus for calcium aluminosilicate (CAS) glasses using NN with 6 hidden layer units.

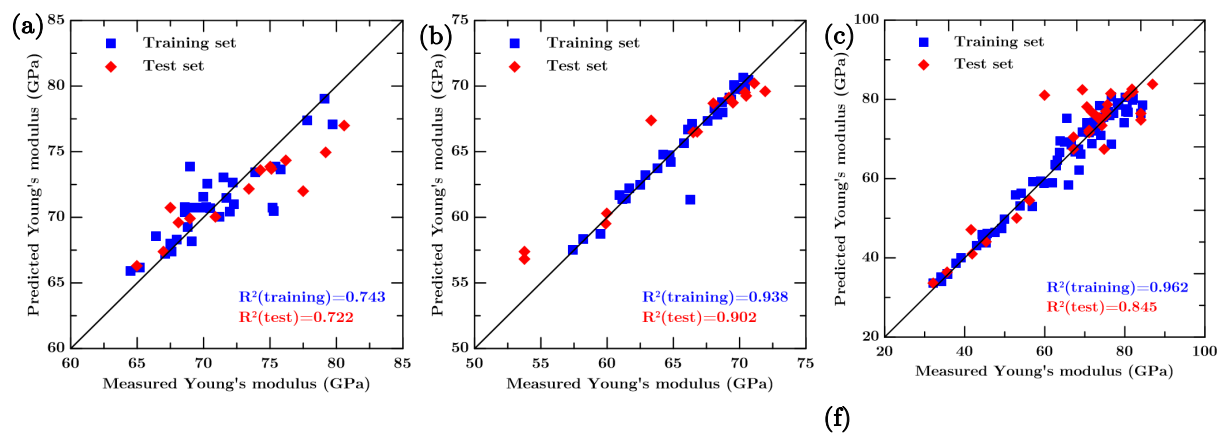

(f)

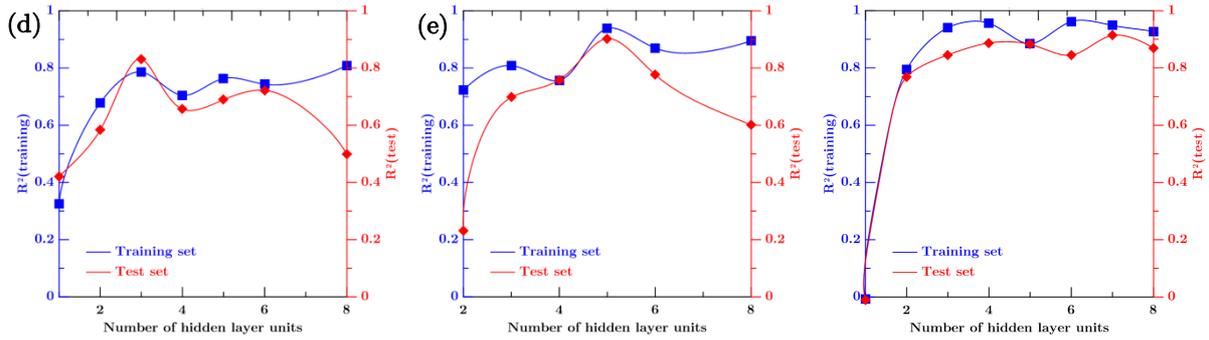

**Figure 2.** Predicted Young's modulus (in GPa) with respect to measured values for (a) sodium calcium silicate (NCS), (b) sodium germanium silicate (NGS) and (c) sodium borosilicate (NBS) glasses corresponding to 6, 5, and 6 hidden layer units, respectively. $R^2$ of training set (left axis) and $R^2$ of the test set (right axis) with respect to the number of hidden layer units in the neural network for (a) NCS, (b) NGS and (c) NBS glasses. Lines are guide for eye.

**Gaussian Process Regression**

Now, we focus on the predictions based on GPR (see Methodology). Figure 3(a) shows the predicted values of Young's modulus of CAS glasses using GPR with exponential kernel functions in comparison to the experimental results. We observe that $R^2$ values of 0.818 and 0.794 are obtained corresponding to the training and test sets, respectively. These results confirm that the GPR exhibits a better agreement with the experiments than NN. Further, $R^2$ values of the test set are comparable to that of the training set suggesting an optimum training. This could be attributed to the nonparametric nature of GPR, which prevents any overfitting of the data. Overall, we observe that GPR can perform better than NN in predicting the properties even for a small dataset.

While the predictions using GPR with exponential kernel function are better than those offered by NN, it still exhibits a notable scatter as exemplified by the $R^2$ value ~ 0.8. In order to improve the predictions, we train the CAS glass data using GPR with automatic relevance determination (ARD) exponential kernel function (see Methodology). Note that ARD kernel functions have an additional degree of freedom through a length scale parameter, which can be tuned to obtain higher predictive accuracy by reducing the noise standard deviation (see Methodology). Figure 3(b) shows the predicted values of Young's modulus of CAS glasses using GPR with ARD exponential kernel functions in comparison to the experimental results. Interestingly, we observe that the predictions using ARD kernel exhibits notable improvement in comparison to exponential kernel function (see Fig. 3(a)) as evidenced by the $R^2$ values of 0.925 and 0.878 for the training and test set, respectively. This could be attributed to the variable length scale parameter employed in ARD kernels due to which the predictions can be improved by assigning differential relevance to the points in the dataset. Overall, these results suggest that GPR with ARD kernels can significantly outperform NN, especially in the case of small datasets.

(b)

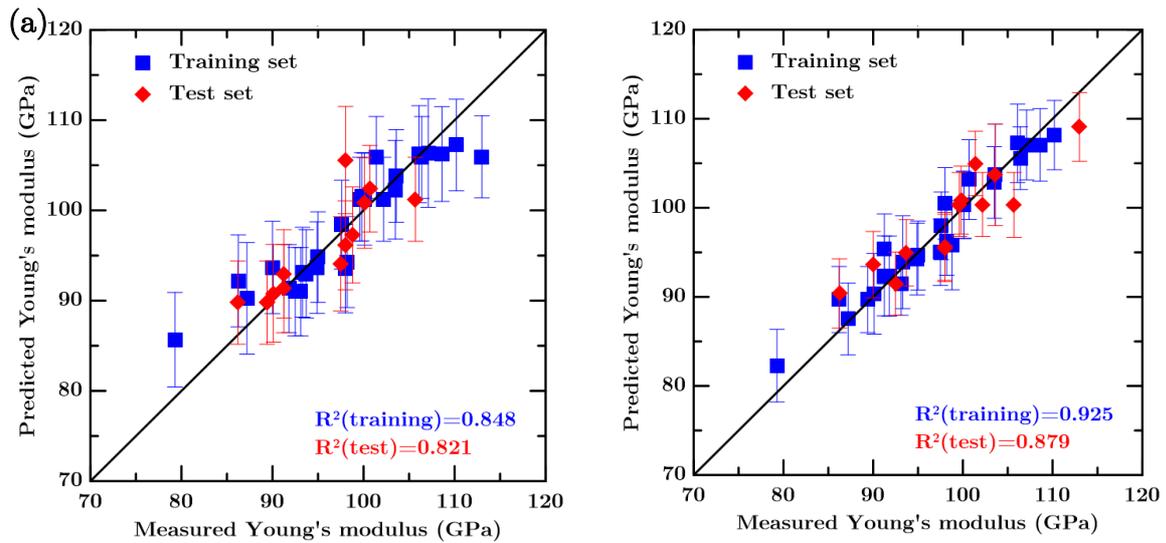

**Figure 3.** Predicted Young's modulus (in GPa) of calcium aluminosilicate(CAS) using the GPR with respect to the measured values, taking kernel function as **(a)** exponential, **(b)** ARD exponential. The symbols represent the mean predicted values and error bars represent the standard deviation corresponding to each mean prediction.

Now, we focus on the predictions of the Young's modulus for the three glass compositions using GPR with ARD exponential kernel. Figures 4(a)–(c) shows the predicted Young's modulus for the NCS, NGS, and NBS glasses using GPR in comparison to their experimental values. We observe that the predictions from the GPR exhibits a close match with the experimental values both for the training and test sets. This is confirmed by the high $R^2$ values for each of the glasses. Further, the standard deviation corresponding to each of the predicted values is represented by the error bars. This provides a quantitative measure of the reliability corresponding to each predicted value, thereby improving the confidence in the predictions. Overall, this establishes the robustness and transferability of GPR to predict properties even from a small dataset.

Further, Table 3 shows the standard deviation of the input values and the predicted Young's modulus values obtained from the GPR for different glass compositions. Note that the standard deviation of the training set represents the level of "noise" present in the signals used for training. On the other hand, the standard deviation in the test set correspond to the uncertainty in the prediction given the distribution of the training data. We observe that the standard deviation for the Young's modulus varies for each glass compositions. It is found to be significantly lower for glass compositions having low input "noise", suggesting that the reliability of predictions can be improved by using a consistent and accurate dataset. Even for the high noise values for the input compositions, it should be noted that the standard deviation observed here is indeed low considering the range of the Young's modulus values for the glasses.

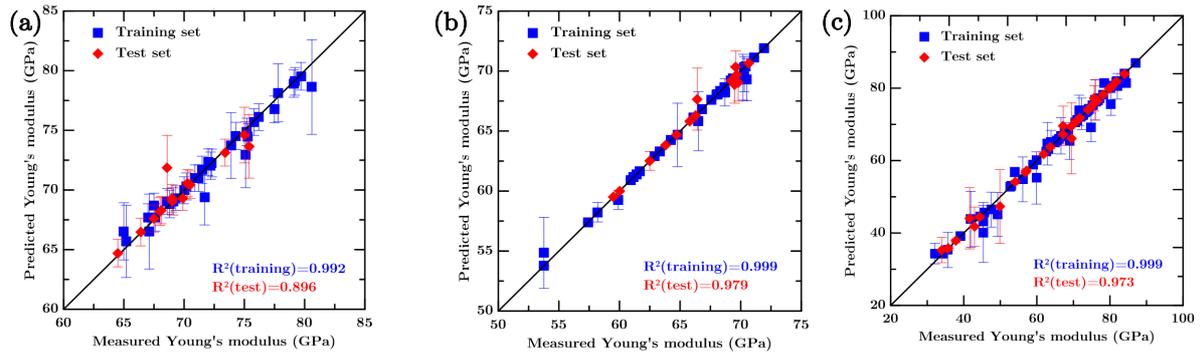

**Figure 4.** Predicted Young's modulus (in GPa) with respect to measured values for (a) sodium calcium silicate (NCS), (b) sodium germanium silicate (NGS), and (c) sodium borosilicate (NBS) glasses using GPR with ARD exponential kernel function, respectively. Note that the symbols represent the mean predicted value and error bars represent the standard deviation corresponding to each predicted value.

**Table 3:** Standard deviation of the Young's modulus values obtained from the GPR with ARD exponential kernel function for different glass compositions considered.

| Glass | Standard deviation of Young's modulus (GPa) in the: | |
|---|---|---|
| | training set | test set |
| CAS | 3.052 | 4.027 |
| NCS | 0.828 | 1.523 |
| NGS | 0.052 | 0.561 |
| NBS | 0.362 | 1.822 |

**DISCUSSION**

Now, we compare the predictive capability of two commonly used methods–MM model (analytical physics-based method), and NN (parametric machine learning method)–with respect to the GPR with ARD exponential kernel functions. Note that the results from NN correspond to that of the optimized NN, that is, ensuring there is no overfitting. Figures 5(a) to 5(d) shows the predicted Young's modulus in comparison to the measured values using MM model, NN, and GPR with ARD exponential function for the four chosen glasses, namely, CAS, NCS, NBS, and NGS, respectively. First, we observe that the machine learning methods, that is, NN and GPR, significantly outperforms the MM model. This is evident from the low $R^2$ values of 0.307, 0.547, and 0.233 for CAS, NCS, and NBS glasses (see Figs. 5(a)–(c)), respectively, for MM model. This is ascribed to the fact that the MM model is essentially a linear additive model wherein stiffness is a linear functional of the composition and density of glasses (see Methodology). As such, MM model is inherently unable to capture any non-linearity in the composition–Young's modulus relationship, which is commonly observed in silicate glasses. While MM model is known to provide a reasonable estimate of the order of magnitude, it systematically under-/over-predicts the slope. Further, it is notoriously unable to predict the Young's modulus of glasses containing borate species[31]. This is due to the differential coordination number boron can take depending on the cations.

Second, GPR clearly outperforms NN for all the glass compositions chosen. This is exemplified by the high $R^2$ values of 0.912, 0.910, 0.981, and 0.999, for CAS, NCS, NBS, and NGS glasses, respectively, predicted using GPR with ARD exponential kernel functions (see Figs. 5(a)–(d)).

This is due to the unique ability of GPR to capture the model uncertainty by identifying the underlying probability distribution from which the data is sampled. Thanks to the nonparametric nature of GPR, it is independent of any structural limitation (hidden layers and number of hidden layer units as in the case of NN) and can be used even for a small dataset to obtain reliable predictions without overfitting. On the contrary, NN is limited by the specific model structure to be learned and hence, can exhibit overfitting[5] depending the data size and the model structure such as number of hidden layers and hidden layer units.

Third, GPR can better capture the data variability due to the dedicated covariance (kernel) function, a functional handle which NN do not have employed in the model. In other words, upon training, GPR identifies the underlying distribution for the dataset wherein the mean corresponds to the predicted value, with the confidence interval of prediction provided by the standard deviation. Thus, GPR enables one to identify whether the predictions from the model are reliable or not in a quantitative manner (see Table 3), thereby addressing one of the fundamental issues in using ML methods for property prediction. This is in contrast with NN, where just one output value is obtained as the prediction. As such, the model uncertainty is not accurately captured by the NN. This makes it challenging to identify whether the predictions from NN for a new data set or an extrapolation are reliable or not.

Finally, using ARD kernel with GPR exhibits improved results with respect to ordinary GPR with exponential function. This is attributed to the nature of the length scale parameter which determine the relevancy of input data to the regression. In the case of GPR with exponential kernel function, the length scale parameter is kept constant. On the other hand, in the case of ARD exponential kernel function, the length scale parameter is treated as a variable. Accordingly, depending on the relevance of the input with respect to the output, the length scale parameter is tuned so as to maximize the predictive capability. For example, if an input is irrelevant for regression output, the length scale parameter is increased so that the kernel functions attains a low value and vice-versa. This is apparent from the $R^2$ values reported herein. GPR model trained with ARD exponential kernel function significantly outperforms the GPR model trained with exponential kernel function.

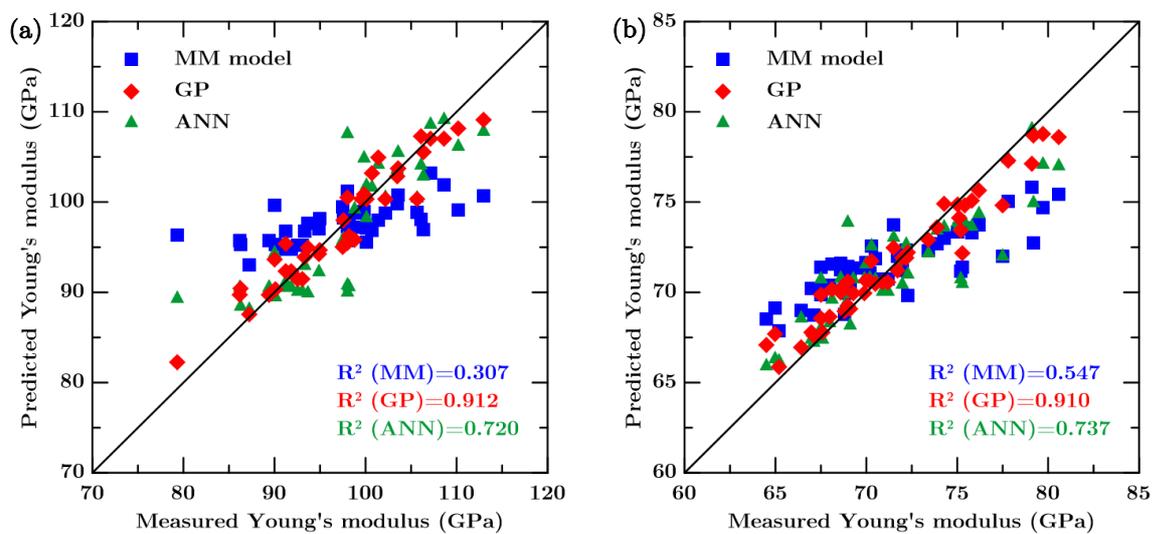

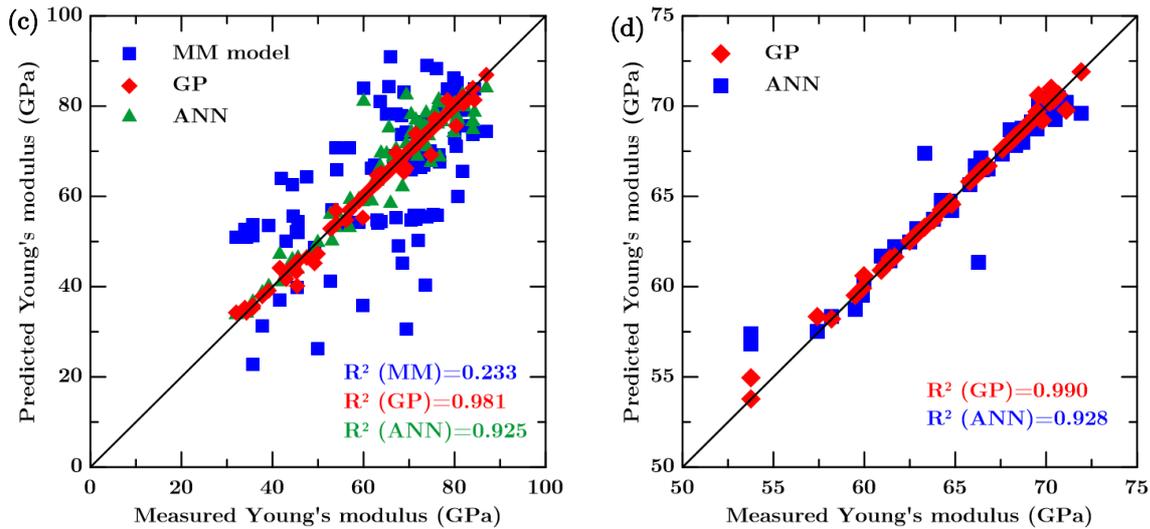

**Figure 5.** Comparing the three different models (MM model, GPR and NN) for (a) calcium aluminosilicate (CAS), (b) sodium calcium silicate (NCS), (c) sodium borosilicate (NBS) glasses, and (d) sodium germanium silicate (NGS). Note that NGS contains only two models (GPR and NN) due to unavailability of data for MM model.

## CONCLUSIONS

Overall, we present a robust methodology using GPR to predict the properties of silicate glasses. We show that GPR provides rigorous estimates for the Young's modulus for a wide range of glass compositions, even for small datasets. Since GPR identifies the underlying distribution corresponding to a dataset, the standard deviation of the distribution obtained can quantitatively provide the reliability of the predictions. Further, thanks to the non-parametric nature of GPR, it avoids overfitting even for a small dataset, which is a fundamental issue observed in NN. Finally, the methodology presented herein is highly transferable and can be used to develop glass compositions with tailored properties, thereby accelerating the development of novel functional glasses.

## ACKNOWLEDGEMENTS

The authors acknowledge the financial support for this research provided by the Department of Science and Technology, India under the INSPIRE faculty scheme (DST/INSPIRE/04/2016/002774). The authors thank the IIT Delhi HPC facility for providing the computational and storage resources.